\documentclass[11pt]{article}
\usepackage{hyperref}
\pdfoutput=1
\begin{document}
\title{Flow Visualisation of Annular Liquid Sheet Instability \& Atomisation}
\author{Daniel Duke, Damon Honnery \& Julio Soria \vspace{6pt}\\
Laboratory for Turbulence Research in Aerospace \& Combustion, \\
Department of Mechanical \& Aerospace Engineering, \\
Monash University, Australia}
\maketitle
\begin{abstract}
Fluid dynamics videos of unstable thin annular liquid sheets are presented in this short paper.  These videos are to be presented in the Gallery of Fluid Motion for the American Physical Society 65th Annual Meeting of the Division of Fluid Dynamics in San Diego, CA, 18-20 November 2012.  An annular sheet of thickness $h=1mm$ and mean radius $R=18.9mm$ is subjected to aerodynamic axial shear from co-flowing air at various shear rates on both the inner and outer surface at a liquid sheet Reynolds Number of $\mathrm{Re}= 500$.
\end{abstract}
\section*{Description of Videos}
A thin annular sheet of thickness $h=1mm$ and mean radius $R=18.9mm$ has been studied using a high-speed video camera and high-speed pulsed light source.   The camera is a PCO DIMAX $2000 \times 2000$ pixel 12-bit CMOS camera and the light source has been custom-built using a large array of consumer-grade 530nm green LEDs driven by a current amplifier at a duty cycle exceeding 100, such that they can be driven past their steady-state maximum safe operating voltage to increase brightness.  The experimental setup and the annular nozzle geometry is described in more detail in \cite{Duke1, Duke2}.    

The annular sheet is exposed to an outer co-flowing gas stream surrounding the sheet and an inner co-flowing air stream encapsulated by the sheet.  This arrangement sets up a Kelvin-Helmholtz instability mechanism which generates a convective unstable instability frequency with a narrow bandwidth \cite{Duke2}.  This instability grows until the sheet breaks into ligaments and droplets, and subsequently the liquid is atomised.  

In this short paper we present some simple flow visualisations to be presented at the American Physical Society's Divison of Fluid Dynamics conference, in the Gallery of Fluid Motion over a range of conditions.  The hyperlinks in the left-hand column of Table \ref{table1} may be followed to obtain high-speed videos stored online.  

The non-dimensional parameters governing the annular flow the thickness-radius ratio $h/R$, the Reynolds Number based on the liquid sheet thickness $h$, density, viscosity and bulk velocity;
\begin{equation}
\mathrm{Re}=\frac{\rho_{\mathrm{liq.}} \overline{U}_{\mathrm{liq.}} h}{\mu_{\mathrm{liq.}}},
\end{equation}
the Weber Number (which includes the surface tension $\sigma$);
\begin{equation}
\mathrm{We}=\frac{\rho_{\mathrm{liq.}} \overline{U}^{2}_{\mathrm{liq.}} h }{\sigma},
\end{equation}
and the inner and outer gas-liquid Momentum Ratios;
\begin{equation}
\mathrm{MR}_{\mathrm{i,o}} = \frac{ \rho_{\mathrm{i,o}} h \overline{U}^{2}_{\mathrm{i,o}} }{\rho_{\mathrm{liq.}} h \overline{U}^{2}_{\mathrm{liq.}}}.
\end{equation}
The shear at the liquid surface may also be described by the velocity deficit $\Delta \overline{U}_{\mathrm{i,o}} = \overline{U}_{\mathrm{i,o}} - \overline{U}_{\mathrm{liq.}}$.

In Table \ref{table1}, only the effect of varying the outer momentum ratio on the flow is demonstrated, due to the large size of the videos.  The Reynolds \& Weber Number are set at $\mathrm{Re}=500,\mathrm{We}=2.5$, $h/R = 5.29 \times 10^{-2}$, and the inner gas flow at $\mathrm{MR}_{\mathrm{i}}=8.759 \times 10^{-1}$, $\Delta\overline{U}_{\mathrm{i}}=9.5$ m/s.  In all cases except Case 1, the gas flows are at a higher speed than the liquid sheet.   Images at a wider range of conditions than those shown here are available in high definition on YouTube: \href{http://youtu.be/XOvk6NumQkw}{\underline{http://youtu.be/XOvk6NumQkw}}.   The images in Table \ref{table1} have been taken at a frame rate of 2500 frames per second and are played back at 25 frames per second, 100 times actual speed.  

\begin{table}
\begin{tabular}{r|l|l|l|l|l}
Case  & $h/R$ & Re, We & $\mathrm{MR}_{\mathrm{i}}$  & $\mathrm{MR}_{\mathrm{o}}$ & $\Delta\overline{U}_{\mathrm{o}}$ \\
\hline \hline
\href{anc/Case1.mp4}{\textbf{Case 1}} & $5.29 \times 10^{-2}$ & 500, 2.5 & 0 & 0 & 0.00 m/s\\
\href{anc/Case2.mp4}{\textbf{Case 2}} & $5.29 \times 10^{-2}$ & 500, 2.5 & 0.8759 & $4.62 \times 10^{-2}$ & 1.85 m/s \\
\href{anc/Case3.mp4}{\textbf{Case 3}} & $5.29 \times 10^{-2}$ & 500, 2.5 & 0.8759 & $3.60 \times 10^{-1}$ & 5.73 m/s \\
\href{anc/Case4.mp4}{\textbf{Case 4}} & $5.29 \times 10^{-2}$ & 500, 2.5 & 0.8759 & $1.08$ & 10.12 m/s\\
\href{anc/Case5.mp4}{\textbf{Case 5}} & $5.29 \times 10^{-2}$ & 500, 2.5 & 0.8759 & $2.16$ & 14.47 m/s\\
\href{anc/Case6.mp4}{\textbf{Case 6}} & $5.29 \times 10^{-2}$ & 500, 2.5 & 0.8759 & $3.63$ & 18.83 m/s\\
\href{anc/Case7.mp4}{\textbf{Case 7}} & $5.29 \times 10^{-2}$ & 500, 2.5 & 0.8759 & $5.50$ & 23.21 m/s\\
\href{anc/Case8.mp4}{\textbf{Case 8}} & $5.29 \times 10^{-2}$ & 500, 2.5 & 0.8759 & $7.71$ & 27.53 m/s\\
\href{anc/Case9.mp4}{\textbf{Case 9}} & $5.29 \times 10^{-2}$ & 500, 2.5 & 0.8759 & $10.35$ & 32.00 m/s\\
\hline
\end{tabular}
\caption{\label{table1}Table of videos for flow visualisation cases.  The effect of increasing the outer gas momentum ratio is demonstrated.  The case name corresponds to the video file on \textit{arXiv}.}
\end{table}

These images have been subsequently analysed using a Fourier image correlation technique \cite{Duke1}, which reveals quantitative data on the surface velocity as a function of both streamwise distance from the nozzle and time.  This time-series data may then be used to obtain frequency information and can also be further processed using Koopman mode decomposition to reveal the streamwise evolution in terms of instability growth rate \cite{Duke2}.  

Hilbert decomposition of the dominant sinusoidal instability has revealed secondary non-linear instability mechanisms due to sheet thinning as a result of the interaction between the faster-moving gas and the liquid sheet which are a key factor in the sheet break-up mechanism \cite{Duke2}.  Over a wide range of conditions beyond those presented in this short paper, the scaling behaviour of both the inner and outer momentum ratio on the frequency of the instability has been determined \cite{Duke3}.  The measurements detailed in the referenced papers are all obtained via digital image processing of the high-speed images using the same techniques as presented here.  In the attached movies, the spatial resolution has been reduced to increase the field of view for an easier interpretation of the global flow-field.

\section*{Copyright Notice}
This article has been published on the \textit{ArXiv} under a perpetual, non-exclusive license.  Copyright is retained by the authors. The attached video files are Copyright (c) 2012 by the authors and may not be copied, publicly presented or incorporated into any other derivative work without a clear attribution and written consent.

\end{document}